\documentclass{aastex631}

\usepackage{amsmath}
\usepackage{savesym}
\savesymbol{tablenum}
\usepackage{siunitx}
\restoresymbol{SIX}{tablenum}

\begin{document}

\title{Integrated Galaxy Light from Stacking {\boldmath $10^5$} Random Pointings in the Dark Energy Survey Data}

\author[0000-0002-7340-9291]{Jenna E. Moore}
\affiliation{Duke University \\
120 Science Drive \\
Durham, NC, 27710 USA}
\affiliation{School of Earth \& Space Exploration, \\
Arizona State University, \\
Tempe, AZ\,85287-6004, USA}

\author[0000-0003-3329-1337]{Seth H. Cohen}
\affiliation{School of Earth \& Space Exploration, \\
Arizona State University, \\
Tempe, AZ\,85287-6004, USA}

\author[0000-0001-6397-5516]{Philip Mauskopf}
\affiliation{School of Earth \& Space Exploration, \\
Arizona State University, \\
Tempe, AZ\,85287-6004, USA}

\author[0000-0002-3193-1196]{Evan Scannapieco}
\affiliation{School of Earth \& Space Exploration, \\
Arizona State University, \\
Tempe, AZ\,85287-6004, USA}

\begin{abstract}

We present a new technique for measuring the integrated galaxy light (IGL) with stacked image data from the Dark Energy Survey (DES). We extract $1\times1$ arcminute cutouts from nearly 100,000 randomly selected positions in the \textit{g, r, i, z,} and \textit{Y} bands from the DES data release 2 (DR2) maps. We generate source catalogs and masks for each cutout and the images are subsequently stacked to generate deep images of the sky both with and without sources. The IGL is then calculated by taking the difference in average brightness between stacks that contain galaxies and stacks in which galaxies have been masked. We find IGL values of $g = 4.27 \pm 0.28, r = 6.97 \pm 0.42, i = 8.66 \pm 0.53, z = 10.16 \pm 0.7,$ and $Y = 13.78 \pm 2.35$ nW/m$^2$/sr. These measurements, which require no foreground estimation or removal, are in agreement with previously reported IGL values derived from galaxy number counts and other methods. This stacking technique reduces the sensitivity to diffuse local backgrounds but is not sensitive to large-scale diffuse extragalactic background light.
\end{abstract}



\section{Introduction} \label{sec:intro}
The extragalactic background light (EBL) is comprised of all of the light emitted over all time by all sources and processes outside of our Milky Way. Following the convention of \cite{cooray16}, the EBL includes all background radiation across the entire electromagnetic spectrum, including that which may be attributed to particular sources (e.g., stars and galaxies) as well as more diffuse signals. The intensity of the EBL has several peaks across the optical, infrared (IR), and microwave regions of the spectrum. These peaks are known as the cosmic microwave (CMB), optical (COB), and infrared (CIB) backgrounds, respectively. The focus of the present work is the COB, which encompasses the optical and near-IR portion of the spectrum from $0.1$ to $5\mu$m. The dominant source of light in this region is stellar nucleosynthesis, making the COB an important metric in constraining models of galaxy formation and evolution \citep{hauser01}. 
 
Unlike the CMB, which can be measured with great precision \citep{planck20}, the COB has proven more difficult to characterize. Direct measurements of the COB via optical/near-IR measurements of the sky are contaminated with a number of foreground signals that must be precisely characterized and removed to reveal the underlying COB signal \citep{dwek98, hauser98, hauser01}. These signals include bright foregrounds from within our solar system (zodiacal light), scattered starlight within our galaxy (diffuse galactic light/DGL), and, for ground-based observations, Earth's atmosphere. For this reason, direct measurements of the COB have typically come from space-based instruments such as the Wide Field and Planetary Camera onboard the Hubble Space Telescope (WFPC2/HST) \citep{bernstein07}, Pioneer 10/11 \citep{matsuoka11}, and the Cosmic Infrared Background Experiment (CIBER) \citep{matsuura17}. In deep space, the Long-Range Reconnaissance Imager (LORRI) onboard the New Horizons spacecraft has provided direct measurements of the COB that are essentially free of zodiacal light contamination \citep{zemcov17, lauer21, symons23, postman24}.  A ground-based measurement method was developed by \cite{mattlia17}, which used Very Large Telescope (VLT) observations of the dark cloud Lynds 1642 to measure the COB at 400 nm by taking the difference in measured brightness between the obscured region and its unobscured surroundings. Although this technique eliminates the need to model and remove zodiacal light, atmospheric airglow, and instrumental noise, the diffuse galactic light scattered by the cloud itself must still be accounted for.  

The integrated galaxy light (IGL) is the sum of all light emitted by resolved galaxies and is expected to be the dominant component of the COB. The IGL is estimated from galaxy number counts on deep survey catalogs such as the Galaxy and Mass Assembly (GAMA) survey and HST archival data, among others \citep{driver16, koushan21}. Measurements of the IGL provide a lower bound on the COB and are limited primarily by the survey depth, as well as the instrument's ability to resolve sources and the faint outskirts of light surrounding them. Moreover, IGL measurements account for the component of the COB attributable to sources, but are not sensitive to diffuse signals. Observations of very high-energy gamma-ray sources (e.g. blazars) serve as an indirect probe of the COB. Photons from these sources are attenuated by their interactions with the EBL as they travel through space, leading to absorption features in their spectra \citep{dwek13}. Historically, COB estimates derived from gamma-ray observations have agreed with the IGL measurements, but with large uncertainties \citep{aharonian06, biteau15, acciari19, desai19}.

While the COB and IGL are expected to be nearly equivalent at optical and infrared wavelengths, there have until recently been significant discrepancies in measured values, with direct COB measurements in the optical regime significantly higher than the values estimated by IGL and gamma-ray studies \citep{driver21}. The tension between direct COB and IGL measurements has been the motivation behind recent analyses such as the SKYSURF project \citep{carleton22, kramer22, obrien23}, which is using archival HST data to better constrain foreground levels and provide upper limits on the EBL at a number of wavelengths. One explanation for the discrepancy may be that direct measurements of the COB have underestimated the amount of foreground contamination (thus resulting in an overestimation of the COB). Alternatively, measurements of the IGL could perhaps have been too low: there may be unresolved sources missing from the galaxy counts, or the faint envelopes of resolved sources may not have been fully accounted for. Notably, \cite{kramer22} investigated the possibility that missing faint galaxies in IGL measurements could explain the discrepancy and found that missing galaxies alone could not explain the difference. This finding aligns with the latest results from New Horizons \citep{postman24} and new, more precise, gamma-ray measurements \citep{greaux24}, both of which agree decisively with the IGL measurements, indicating the COB is very likely entirely made up of IGL. 

In this paper, we present a new technique for estimating the IGL at optical and near-infrared wavelengths using images from the Dark Energy Survey (DES). Borrowing a method commonly employed to extract faint signals in millimeter and sub-millimeter astronomy, we average together (``stack'') images of a large number ($\approx 100,000$) of random positions on the sky. While the IGL is commonly estimated through number counts of galaxies from existing catalog data, we instead use software tools to identify and categorize sources in the images. We then use these catalogs to generate source masks and recover the IGL by taking the difference between stacks containing galaxies and stacks in which detected galaxies have been masked. Since all stacks are comprised of the same set of images, taking the difference between them also subtracts out any foregrounds (e.g. zodiacal light, diffuse galactic light) or instrumental noise. Thus, we have a relatively simple method for measuring the IGL from direct images of the sky taken from the ground that requires minimal corrections. We note that this technique measures the IGL only and is not sensitive to diffuse light. 

\section{Data} \label{sec:data}

The DES Dark Energy Camera (DECam) is an optical/near-infrared camera mounted on the 4-meter Victor M. Blanco telescope located in the Cerro Tololo Inter-American Observatory (CTIO) in Chile. DECam measures five photometric bands: $g = 477, r = 637, i = 777, z = 916, \text{and } Y = 989 \text{ nm}$ \citep{flaugher15}. Data Release 2 (DR2) includes data from six years of observations, with co-add image depths of $g = 24.7$, $r = 24.4$, $i = 23.8$, $z = 23.1$, and $Y = 21.7$ mag (AB) and point-spread function full width at half maximum (FWHM) of $g=1.11''$, $r=0.95''$, $i=0.88''$, $z=0.83''$, and $Y=0.90''$ \citep{des21}. For this analysis, over 100,000 random positions were selected within a 2,500 square degree patch of the southern sky, uniformly distributed in the overlapping region of the DES and South Pole Telescope (SPT) footprints. Separate stacking analyses of quiescent galaxies in this region have been conducted using DES data (to be presented in another paper) as well as with data from SPT and the Atacama Cosmology Telescope (ACT) to study the thermal Sunyaev-Zel'dovich effect \citep{meinke21, meinke23}. The use of such a large number of random pointings minimizes possible effects of cosmic variance \citep[e.g.][]{driver10}.  
 
The DES cutout server was used to request cutouts from the DR2 co-add image tiles. For each random position and each of the five DES bands, $1\times1$ arcminute ($228\times228$ pixels) cutouts centered on the coordinates were downloaded from the cutout server to the Sol supercomputer at Arizona State University \citep{jennewein23} for processing and stacking. Each cutout was obtained in the form of a multi-extension FITS file containing science, weight (inverse variance), and mask frames. The images have a pixel scale of $0.2637''/\text{pix}$ and an AB zero-point magnitude of 30.0 mag \citep{morganson18}. With each position having five such cutouts (1 per DES band), over 500,000 total cutouts were obtained for this analysis. 

\section{Methods} \label{sec:methods}
\subsection{Image Processing} \label{sec:imgproc}
\begin{figure*}[ht]
\centering
\includegraphics[width=0.75\textwidth]{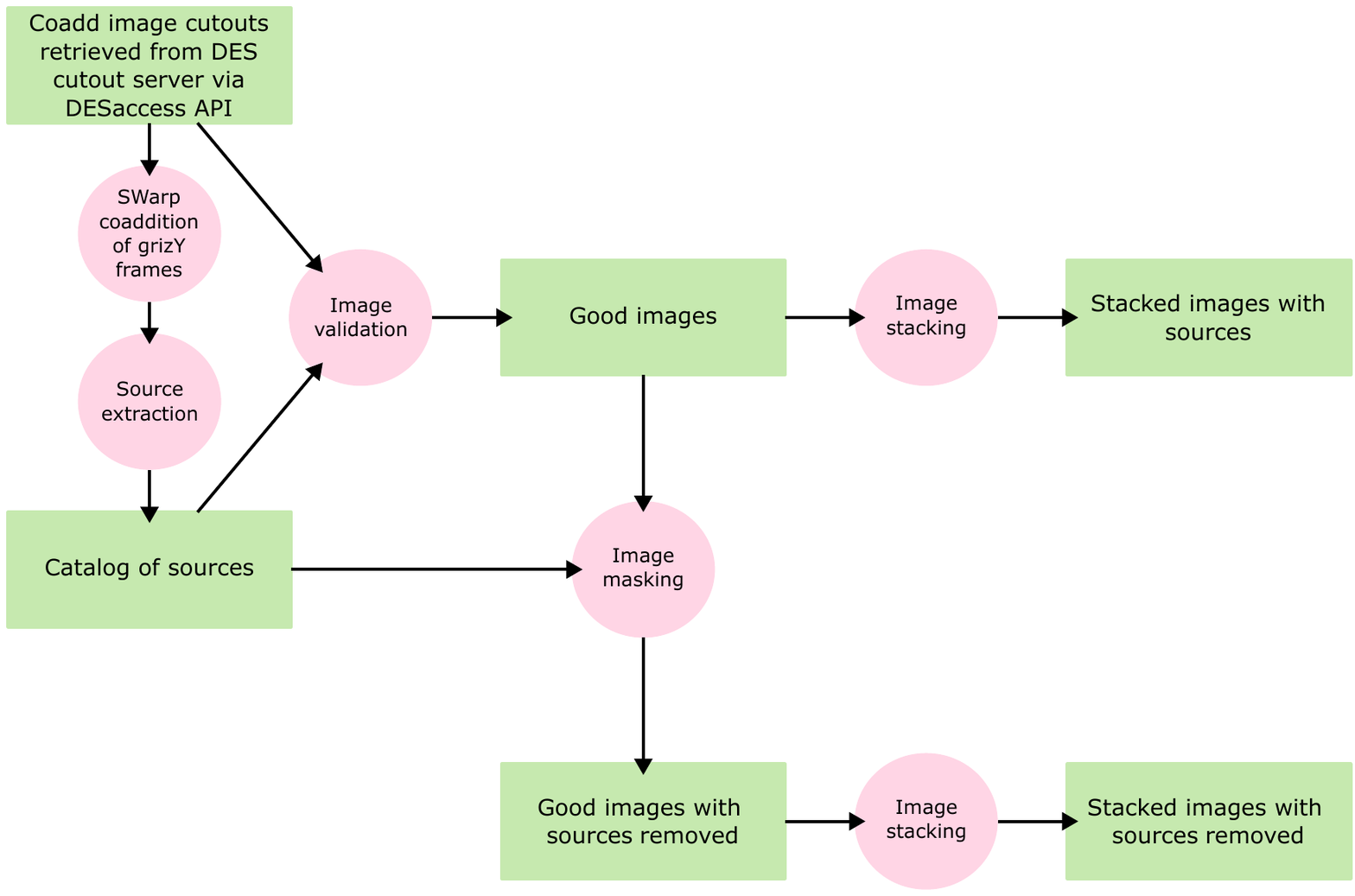}
\caption{Flow chart of the image processing pipeline. Green rectangles represent data products while pink circles represent processing steps. Raw cutouts from the DES cutout server are first co-added to generate detection images, which are used to identify sources. A series of cuts are applied to eliminate images not suited for stacking. Source masks are generated for the surviving cutouts and the images are stacked in a myriad of ways, as described in Section \ref{sec:stacking}.
\label{fig:pipeline}}
\end{figure*}

In this section, we describe the image processing pipeline (Figure \ref{fig:pipeline}), which uses a combination of well-established astronomical software tools and newly developed \texttt{Python} routines. Due to the sheer number of images, we parallelized all image processing and stacking processes to speed up the overall processing time. 

The first step in the pipeline was to generate a detection image for source identification purposes. We accomplish this with two popular astronomical software packages: \texttt{SWarp} \citep{bertin02} and \texttt{SourceExtractor} \citep{bertin96}. \texttt{SWarp} performs re-projection and co-addition of FITS images. \texttt{SourceExtractor} performs source detection on astronomical images and generates object catalogs. We first used \texttt{SWarp} to co-add all five images for each position into a single ``detection'' image. We then fed the detection image into \texttt{SourceExtractor}  along with each individual science and weight frame to detect sources and generate source catalogs and segmentation images for each image. Using \texttt{SourceExtractor} in this dual image mode allowed us to compute source photometry in each band while ensuring that the same sources in a given image are identified in all bands. 

Following source identification, we carried out a series of quality checks to identify and exclude subpar images from the rest of the processing pipeline. We made cuts to exclude images that were saturated by bright sources, contaminated with spurious signals, or that contained other flags indicating poor image quality. We also discarded images that were not $228 \times 228$ pixels (e.g., locations near the edge of a DES tile) to avoid having to re-align images of varying dimensions. To keep the number of images within each stack consistent, we removed all cutouts deemed unsuitable for any reason in any band. After these cuts, the total number of positions suitable for stacking was 98,713. 

\subsection{Source Classification and Masking} \label{sec:sources}
\begin{figure*}[ht]
\includegraphics[width=\textwidth]{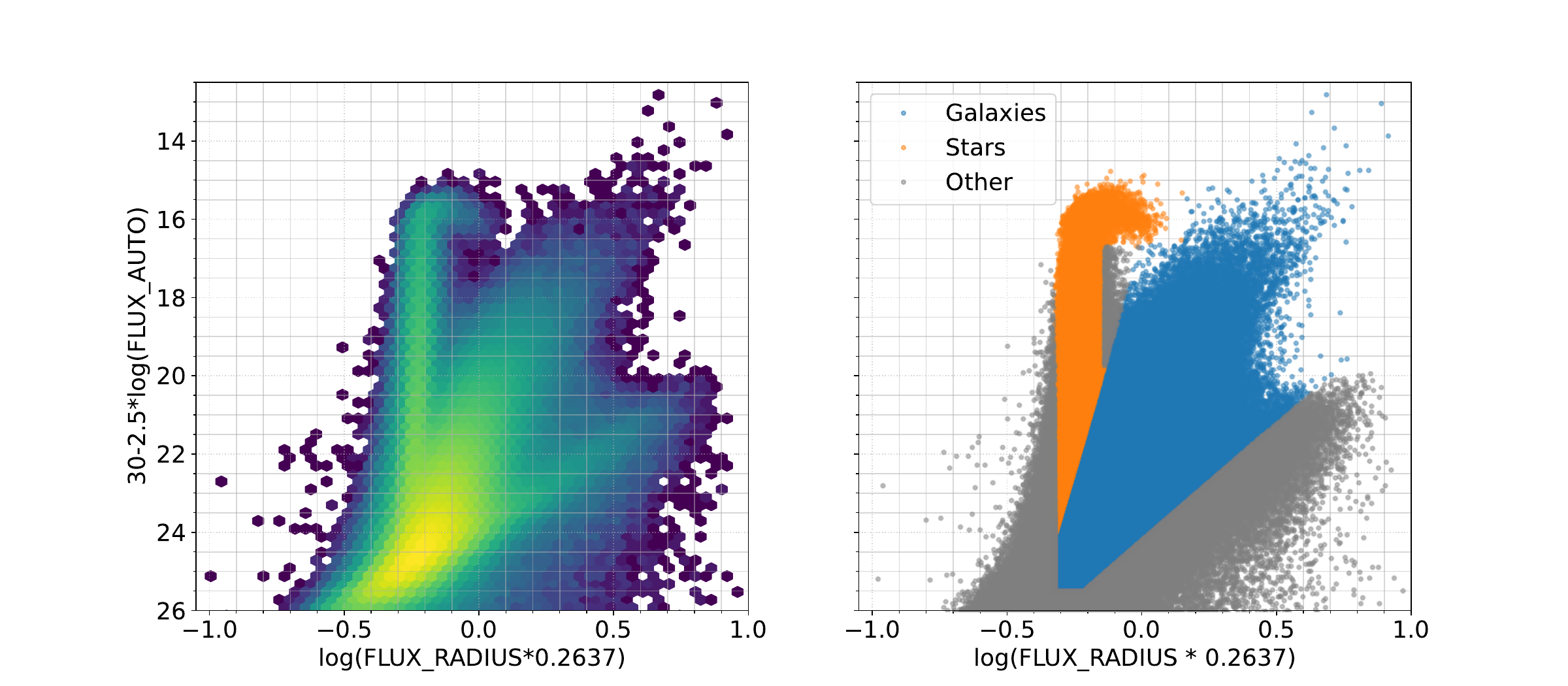}
\caption{r-band magnitude (in AB mag) versus size (in arcsec) distribution for \texttt{SourceExtractor}-detected sources in all 98,713 cutout positions. We used the hexbin plot on the left to generate cuts to separate stars and galaxies. The scatter plot on the right shows the resulting source classification, with gray dots representing the ``other'' sources.
\label{fig:sources}}
\end{figure*}

For the remaining 98,713 positions, we used \texttt{SourceExtractor} r-band photometry to sort detected sources by size and magnitude into three categories: galaxies, stars, and ``other'' discarded sources --- that is, sources that did not clearly fall into either category. Figure \ref{fig:sources} shows the magnitude-size distribution for all identified sources (3,346,610 total) in all 98,713 frames as well as their subsequent classification. 

To achieve this classification, we first cut along the bottom edge of the clearly defined region of galaxies and discarded all sources fainter than 25.5 mag as well as the faint sources that fell below the line
\begin{equation}
    \texttt{mag}= -6\log{[\texttt{FLUX\_RADIUS}}] + 24.2,
\end{equation}
where $\texttt{mag} = 30.0-2.5\log{[\texttt{FLUX\_AUTO}]}$ and \texttt{FLUX\_RADIUS} has been multiplied by the pixel scale ($0.2637''/\text{pix}$). Here, \texttt{FLUX\_AUTO} is the \texttt{SourceExtractor}-computed flux within a Kron aperture and \texttt{FLUX\_RADIUS} is the \texttt{SourceExtractor}-computed half-light radius. We also discarded any sources with $\log{[\texttt{FLUX\_RADIUS}]} < -0.325$. We then identified and discarded the 884 sources that fall in between the clearly defined populations of stars and galaxies. The discarded sources were grouped into the third category of ``other'' sources, of which there were 485,965 total. These sources are likely a mix of true sources and image artifacts. Of the 3,346,610 total detected sources, 190,429 were classified as stars and 2,670,216 were classified as galaxies. 

\begin{figure*}[ht!]
\centering
\includegraphics[height=\textheight]{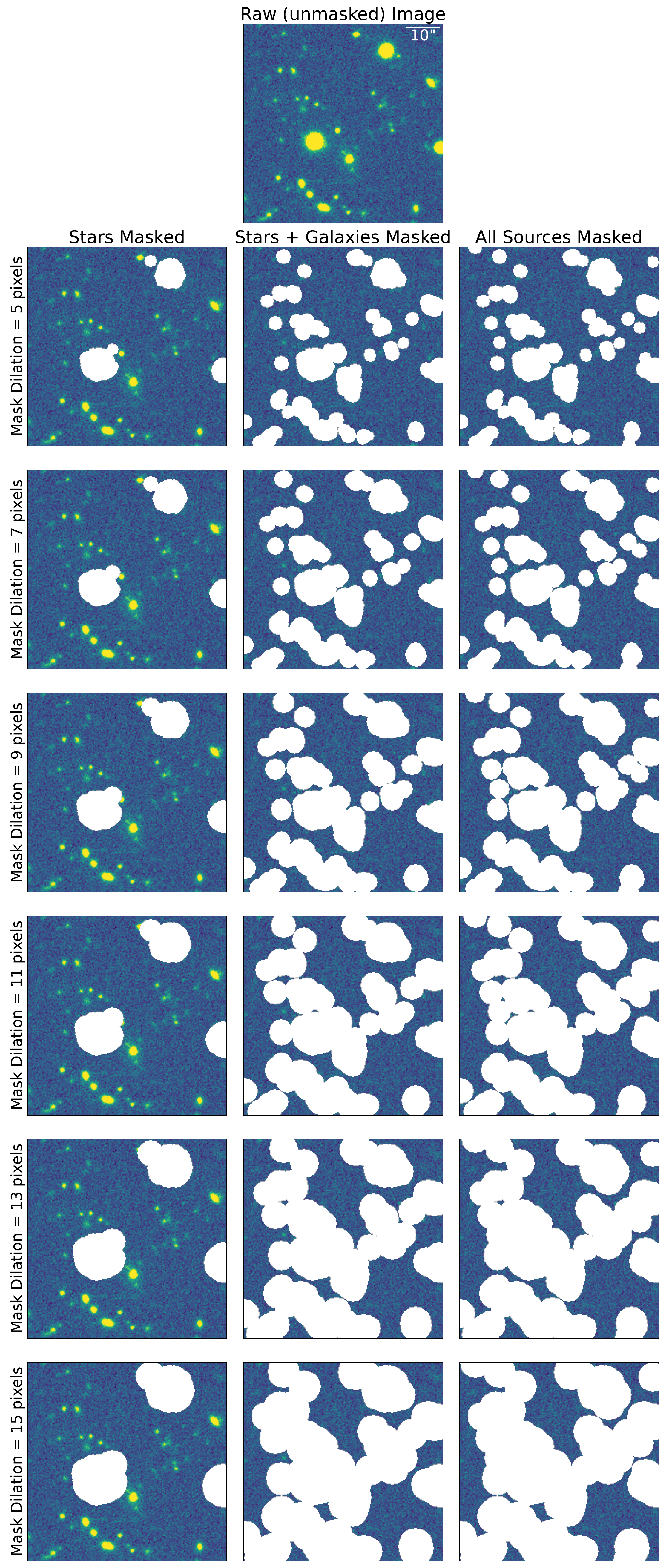}
\caption{An example set of masked images created for one r-band cutout. This same set of images is created for each image for all five bands. Images are $1\times 1$ arcminute squares. 
\label{fig:indimageset}}
\end{figure*}

We used the \texttt{Photutils} \citep{bradley22} \texttt{segmentation} module to generate masked images for three types of sources from the \texttt{SourceExtractor} segmentation image data for each cutout. In the first stacked image, only the sources classified as stars were masked. The second image masked both stars and galaxies, while the third masked \textit{all} \texttt{SourceExtractor}-identified sources. In each image, we applied circular masks around the selected sources. We generated six variations of each masked image, with each variation using a different source mask radius (5, 7, 9, 11, 13, or 15 pixels) to the circular source masks. The dependence of the resulting stacked flux on mask dilation factor (the radius of the circular mask) allows us to statistically fit a smooth model for the dependence of the masked flux on radius independent of the individual galaxy profiles.
In total, for each cutout, there were 18 different masked images for each band. An example image set is shown in Figure \ref{fig:indimageset}

\subsection{Image Stacking} \label{sec:stacking}
\begin{figure*}[ht!]
\includegraphics[width=\textwidth]{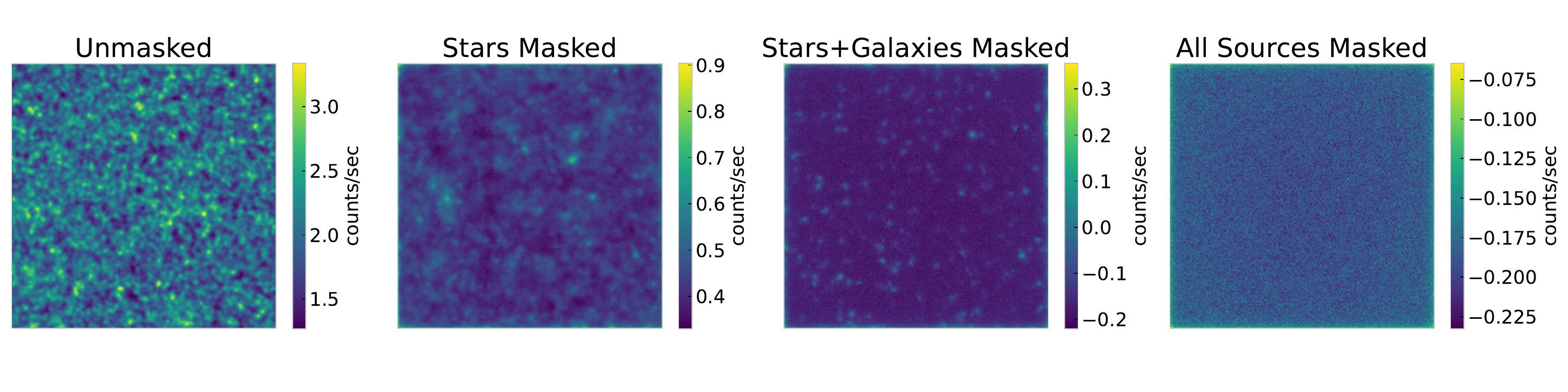}
\caption{Stacked r-band images generated from the weighted average of 98,713 individual cutouts. In this series of masked images, a 15-pixel dilation factor was applied to the masks for each source. Images are $1\times 1$ arcminute squares.
\label{fig:imageset}}
\end{figure*}

The image stacking process is straightforward. We created stacked images by aggregating all the science and weight frames for each of the 18 masked images into $228\times228\times98,713$ arrays and computing the weighted average of each pixel along the third axis. We set masked pixels to \texttt{NaN} and assigned zero weight. We similarly generated ``Unmasked'' stacked images from the raw DES cutouts. An example set of stacked images (generated using a source mask dilation factor of 15 pixels) is shown in Figure \ref{fig:imageset}.

As discussed in Section \ref{sec:sources}, there are four distinct types of stacked images in each of the five bands, each comprised of light from a different mix of sources. All images contain unwanted signals from foregrounds (i.e., zodiacal light, diffuse galactic light) as well as instrumental noise. The four types of stacks differ in which types of sources are masked within them. The unmasked stacked images, by definition, include all sources: galaxies, stars, and the ``other'' sources described in Section \ref{sec:sources}. The stars-masked stacked images contain light from galaxies and other sources, and the stars-and-galaxies-masked stacked images contain light only from the other unclassified sources. We removed these sources in the final all-sources-masked stacked images, which contain just the various sources of noise. 

\begin{figure*}[ht]
\includegraphics[width=\textwidth]{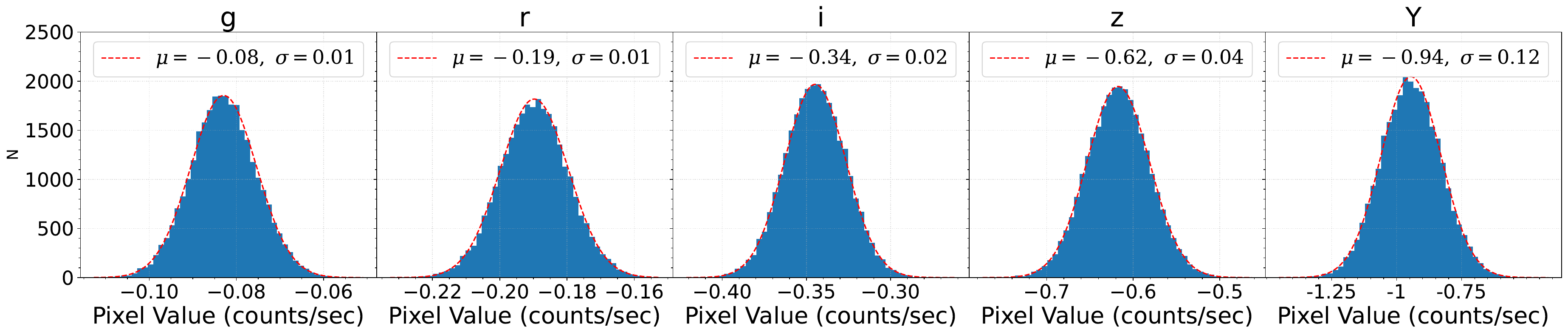}
\caption{Pixel distributions for stacked images with all sources masked (source mask dilation factor = 15 pixels), calculated within a 25 arcsecond circular aperture shown in Figure \ref{fig:avg_aperture}. In all bands, the distributions fall below zero, suggesting a possible overestimation of the background in the DES pipeline.
\label{fig:all_masked_hist}}
\end{figure*}

Of note, all of the stacked images in which all detected sources are masked have all negative pixel values (Fig. \ref{fig:all_masked_hist}). While we did not perform background subtraction on the images as part of this analysis, the DES DR2 cutouts themselves are background subtracted as part of the DES image processing pipeline \citep{morganson18}. Theoretically, with all of the sources removed, these stacks should contain only noise. This appears to be the case, as the distributions in Figure \ref{fig:all_masked_hist} are all Gaussian. However, the negative pixel values suggest that the background may have been overestimated. As we discuss in Section \ref{sec:analysis}, this background overestimation does not affect our brightness calculations since we take the difference between stacks, which naturally subtracts out the signals common to both images (i.e., the noise).

\section{Analysis} \label{sec:analysis}

\subsection{Brightness Measurement}

\begin{figure}[ht!]
\centering
\includegraphics[width=0.5\textwidth]{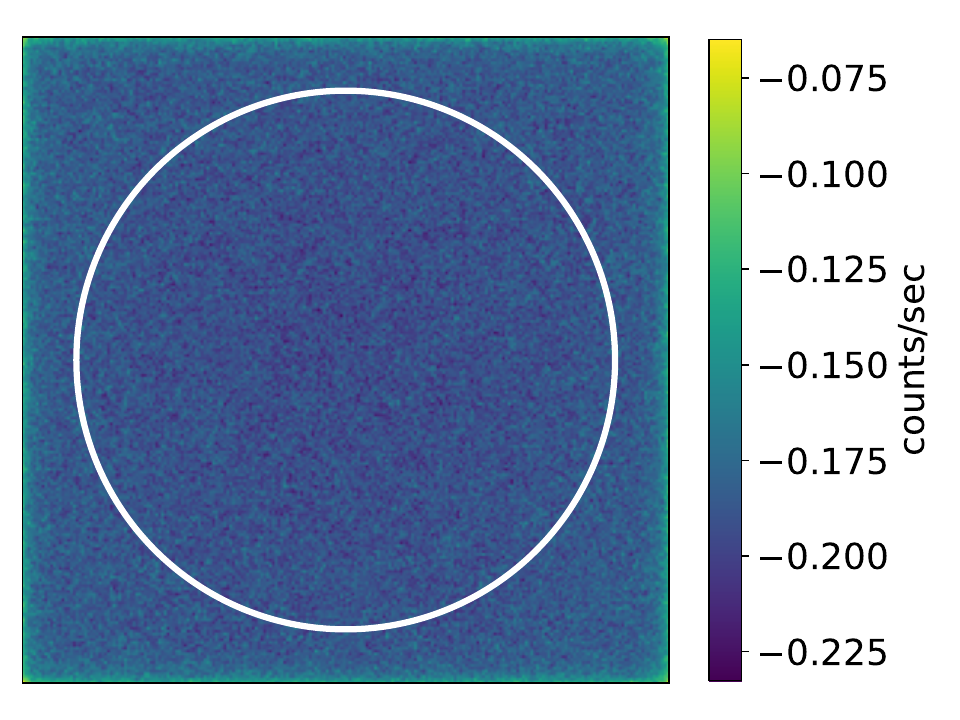}
\caption{Stacked r-band image with all detected sources masked (source mask dilation factor = 15 pixels). The white circular aperture (radius $\approx25$'') encloses the pixels used in the brightness calculation. Edge pixels, which are slightly brighter due to undetected sources, are excluded.
\label{fig:avg_aperture}}
\end{figure}

 We used aperture photometry to calculate the average brightness of each stacked image. We computed the average brightness within a central circular aperture with a 95 pixel ($\approx25''$) radius (Fig. \ref{fig:avg_aperture}). Use of an aperture slightly smaller than the image prevents the average from being skewed by the outer pixels, which are slightly brighter due to undetected sources that lie on or just beyond the edges of individual cutouts. 
 
 For each source mask dilation factor $R$, we calculate the brightness of each of the three types of sources using the average brightnesses $I$ measured from each of the different stacked images as:
 \begin{eqnarray}
     I_{*}(R) &=& I_{\rm unmasked} - I_{*\; \rm masked},\\
     I_{\rm G}(R) &=& I_{*\; \rm masked}- I_{* \rm G\; \rm masked}, \,\,\, {\rm and} \\
    I_{\rm other}(R)&=& I_{*G\; \rm masked} - I_{\rm all\; \rm masked}. 
\end{eqnarray}
 Here, $I_*$ is the brightness from stars, calculated from the difference between the unmasked and stars-masked stacks, $I_{\rm G}$ is the brightness from galaxies, calculated from the difference between the stars-masked and stars-and-galaxies-masked stacks, and $I_{\rm other}$ is the brightness from the unclassified sources, calculated from the difference between the stars-and-galaxies-masked and all-sources-masked stacks. We measured the total brightness $I_{\rm total}$ from the unmasked stacked image, normalized by the brightness of the all-sources-masked stacked image to remove unwanted noise. Using the DES zeropoint magnitude (30.0 mag) and pixel scale ($0.2637''$/pix), we converted the resulting brightness values from image units (counts/second) to surface brightness (SB) units (mag/arcsec$^2$) according to:
 \begin{equation}\label{eq:sb}
     \text{SB} = 30 - 2.5\log\left(\frac{{\rm counts}/{\rm sec}}{(0.2637''/{\rm pix})^2}\right).
 \end{equation}
We then converted these values from surface brightness to MJy/steradian as 
\begin{equation}
    \text{MJy/sr} = 10^{\frac{\text{mag}-8.9}{-2.5}}\times (206265''/\text{rad})^2 \times 10^{-6},
\end{equation}
and finally, to conventional IGL units of $\nu I_{\nu}$ as
\begin{equation}
    \nu I_{\nu}\; ({\rm nW/m^2/sr}) = \text{MJy/sr} \times 10^{-11} \times \frac{c}{\lambda},
\end{equation}
where $c$ is the speed of light and $\lambda$ is the pivot wavelength for the band.

Taking the difference in measured brightness between the different types of image stacks (which are all comprised of the same 98,713 images, with different pixels masked) allows us to recover the brightness associated with each type of source while subtracting out the light associated with foregrounds and instrumental noise that are common to all stacks. The brightness values calculated for each group of sources for each source mask size are presented in Table \ref{tab:brightness}. Here we will focus on the light associated with galaxies, and how we extrapolate these measurements to estimate the IGL.

\begin{table}[h]
\centering
    \begin{tabular}{|c|c|c|c|c|c|}
    \hline
    Band & Mask Dilation Factor (pix) & Total Brightness & Stars & Galaxies & Other \\ \hline
\textbf{g} & 5 & 17.92 & 13.84 & 3.92 & 0.17 \\
 & 7 & 18.02 & 13.85 & 3.99 & 0.18 \\
 & 9 & 18.09 & 13.87 & 4.04 & 0.18 \\
 & 11 & 18.13 & 13.87 & 4.07 & 0.18 \\
 & 13 & 18.16 & 13.88 & 4.09 & 0.19 \\
 & 15 & 18.18 & 13.88 & 4.11 & 0.19 \\ \hline
\textbf{r} & 5 & 23.95 & 17.51 & 6.24 & 0.21 \\
 & 7 & 24.08 & 17.53 & 6.33 & 0.22 \\
 & 9 & 24.17 & 17.55 & 6.39 & 0.22 \\
 & 11 & 24.23 & 17.56 & 6.44 & 0.23 \\
 & 13 & 24.27 & 17.57 & 6.47 & 0.22 \\
 & 15 & 24.3 & 17.58 & 6.5 & 0.22 \\ \hline
\textbf{i} & 5 & 26.29 & 18.3 & 7.71 & 0.29 \\
 & 7 & 26.44 & 18.33 & 7.81 & 0.3 \\
 & 9 & 26.55 & 18.36 & 7.89 & 0.31 \\
 & 11 & 26.62 & 18.38 & 7.94 & 0.31 \\
 & 13 & 26.67 & 18.39 & 7.98 & 0.31 \\
 & 15 & 26.71 & 18.4 & 8.01 & 0.31 \\ \hline
\textbf{z} & 5 & 28.34 & 19.34 & 8.64 & 0.36 \\
 & 7 & 28.52 & 19.39 & 8.76 & 0.37 \\
 & 9 & 28.66 & 19.43 & 8.85 & 0.38 \\
 & 11 & 28.76 & 19.46 & 8.92 & 0.38 \\
 & 13 & 28.83 & 19.48 & 8.97 & 0.38 \\
 & 15 & 28.88 & 19.49 & 9.01 & 0.38 \\ \hline
\textbf{Y} & 5 & 30.42 & 21.27 & 8.76 & 0.38 \\
 & 7 & 30.72 & 21.37 & 8.94 & 0.4 \\
 & 9 & 30.97 & 21.47 & 9.09 & 0.42 \\
 & 11 & 31.16 & 21.55 & 9.19 & 0.43 \\
 & 13 & 31.31 & 21.61 & 9.27 & 0.43 \\
 & 15 & 31.44 & 21.66 & 9.35 & 0.43 \\ \hline
    \end{tabular}
    \caption{Brightness levels measured via aperture photometry of stacked images with varied source mask dilation factors. All brightness values are reported in ${\rm nW/m^2/sr}$. A larger dilation factor indicates a larger source mask. Total brightness is measured from the raw stacked images (no source masking applied) and is normalized by the average brightness of the image with all sources masked. The contribution from stars was calculated by taking the difference of the total brightness and the brightness calculated from the stacked image with all stars masked. Likewise, the contribution from galaxies was calculated by taking the difference of the stacked image with all stars masked and the stacked image with all stars and galaxies masked. Finally, the contribution from other (unclassified) sources was calculated by taking the difference of the stacked image with stars and galaxies masked and the stacked image with all detected sources masked.
    \label{tab:brightness}}
\end{table}

\subsection{Reddening Correction}

Dust within our Milky Way galaxy has a dimming effect on extragalactic sources, which can significantly affect the calculation of source magnitudes. This is especially important for galaxies, as dust shrouds their faint outer envelopes and makes them appear smaller than they truly are. Although our technique of applying source masks of a range of sizes accounts for this particular effect, we must still account for the overall dimming of light from all sources due to dust. To achieve this, we calculated and applied a magnitude correction to the brightness measured for each stacked image. We used the \texttt{gdpyc} Python package to obtain $E(B-V)$ values for each catalog position from the Schlegel, Finkbeiner, and Davis dust maps \citep{schlegel24}. The $E(B-V)$ values generally fall below 0.05, with approximately 10\% of positions having higher value, with a maximum $E(B-V)$ of 0.12. The $E(B-V)$ distribution has a mean of 0.024 and a sigma-clipped standard deviation of 0.008. We used the mean $E(B-V)$ value and used $R_b$ values from \cite{abbott18} to obtain an average magnitude correction $A_b$ for each band according to $A_B = E(B-V) R_B,$ which was subsequently subtracted from the values obtained in Equation \ref{eq:sb}.
 
\subsection{IGL Calculation} \label{sec:igl}
\begin{figure*}[h!]
\includegraphics[width=\textwidth]{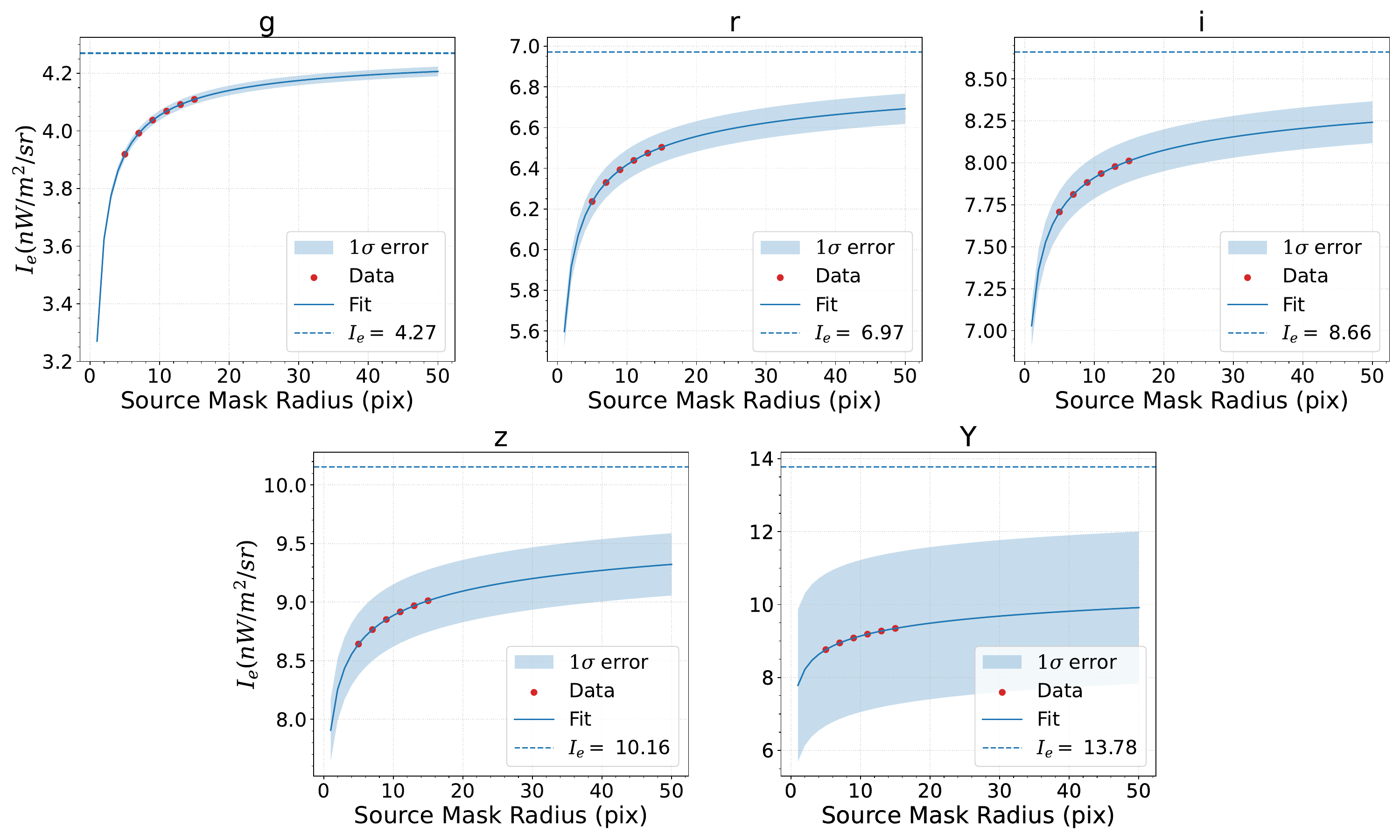}
\caption{Extrapolation of IGL values (dashed lines) from measured galaxy brightness (red points) as a function of source mask dilation size. The shaded blue region represents $1\sigma$ fit error. Larger \textit{Y}-band errors may stem from a shorter exposure time compared to \textit{griz} bands.
\label{fig:igl_fit}}
\end{figure*}

A concern with measuring the background level of an image in which sources have been masked is that the level may be skewed high if the source masks are not sufficiently large. However, to obtain a robust measurement we would like to maximize the number of pixels with which we calculate the brightness. We address these competing concerns by generating a series of stacked images over a range of source mask dilation factor sizes to see how the measured galaxy brightness $I_G$ changes as a function of source mask radius. Figure \ref{fig:igl_fit} shows that in all bands, the measured $I_G$ increases sharply with source mask dilation factor size $R$ below $\approx 10$ pixels before leveling off at larger mask sizes. To obtain the IGL measurement for each band, we fit an empirically adopted function inspired by the Sérsic profile,
\begin{equation}\label{eq:igl_fit}
    I_G(R) = I_e[1-\exp^{-7.669[(\frac{R}{R_e})^{1/n}-1]}],
\end{equation}
to the $I_G$ versus $R$ plot and fit for the IGL ($I_e$), the radius $r_e$, and the factor $n$. The full set of fit parameters and their associated errors are presented in Table \ref{tab:fit_params}. We note the error in the \textit{Y}-band fit is much larger than the other bands, which may be at least partially explained by the much shorter exposure time (450 seconds) for \textit{Y}-band images compared to the other bands (900 seconds) \citep{morganson18} or possibly due to known variable night-sky emission lines at these wavelengths.

\begin{table*}[h]
    \centering
    \begin{tabular}{|c|c|c|c|}
    \hline
         Band & $R_e$ (error) & $I_e$ (error) & $n$ (error) \\ \hline
         g & \num{7.67e-2} (\num{1.55e-2}) & 4.27 (\num{1.68e-2}) & 14.81 (\num{9.31e-1}) \\
         r & \num{8.68e-3} (\num{4.40e-3}) & 6.97 (\num{7.45e-2}) & 24.71 (2.84) \\
         i & \num{3.46e-3} (\num{2.45e-3}) & 8.66 (\num{1.25e-1}) & 28.77 (4.13) \\
         z & \num{1.08e-3} (\num{1.03e-3}) & 10.16 (\num{2.66e-1}) & 38.07 (6.99) \\
         Y & \num{3.37e-4} (\num{5.28e-4}) & 13.78 (2.08) & 77.58 (31.72) \\ \hline
    \end{tabular}
    \caption{IGL fit parameters and errors derived from fitting Equation \ref{eq:igl_fit} to measured galaxy brightness over the range of source mask dilation factors. $R_e$ values are presented in pixels. $I_e$ values are presented in ${\rm nW/m^2/sr}$. $n$ is a dimensionless parameter.}
    \label{tab:fit_params}
\end{table*}

\subsection{Jackknife Tests} \label{sec:jackknife}
\begin{figure*}[ht!]
\includegraphics[width=\textwidth]{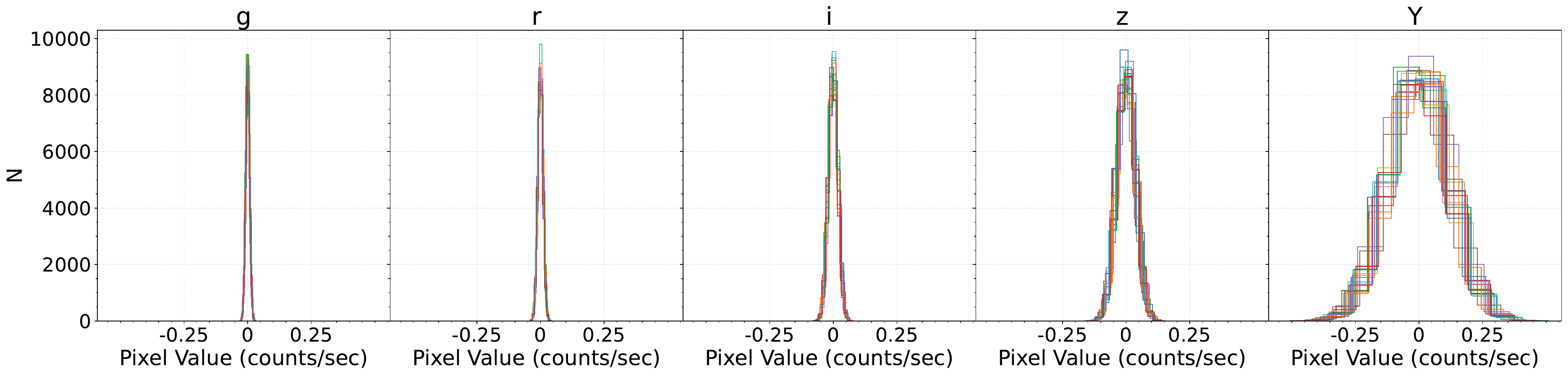}
\caption{Distributions of pixel values in sixteen sets of jackknife images. 
\label{fig:jackknife}}
\end{figure*}

To ensure the robustness of the stacking procedure, we performed a series of jackknife tests on the images. In this  analysis, sixteen different instances of the stacked image with all sources masked with a 15-pixel source mask dilation factor were generated. In each iteration of the jackknife stack, a random $50\%$ of the images were multiplied by -1. The images were then stacked according to the method described in Section \ref{sec:stacking}. Applying the same circular aperture used in the analysis of the real stacked images, histograms of the jackknife images (Figure \ref{fig:jackknife}) are Gaussians well-centered around zero, as expected.

\subsection{Error Estimation}
We calculate the error on each IGL measurement as a quadrature sum of several terms
\begin{equation}\label{eq:error}
    E_{\rm tot} = \sqrt{E_{*}^2 + E_{*\rm G}^2 + E_{\rm fit}^2 + E_{\rm jack}^2},
\end{equation}
where $E_*$ and $E_{*\rm G}$ are the errors derived from the images used in the difference calculation (stars-masked and stars-and-galaxies-masked), $E_{\rm fit}$ is the error associated with the fit of Equation \ref{eq:igl_fit} to the galaxy brightness data, and $E_{\rm jack}$ is the error derived from the jackknife analysis. To calculate $E_*$ and $E_{*\rm G}$, we computed the standard deviation inside the circular aperture for each iteration of the image (e.g., for all source dilation factors), then calculated the average of the standard deviations. Conversely, we take the error from the jackknife analysis to be the standard deviation of the average brightnesses calculated in the circular aperture in each of the sixteen jackknife images. These error values are presented alongside the IGL values in Table \ref{tab:igl}.

\begin{table*}[h!]
    \centering
    \begin{tabular}{|c|c|c|c|c|c|c|c|}
    \hline
        Band & Wavelength (nm) & IGL & $E_*$ & $E_{*G}$ & $E_{fit} $ & $E_{jack}$ & $E_{total}$ \\ \hline
        g & 477.08 & 4.27 & 0.23 & 0.16 & 0.02 & 0.003 & 0.28 \\
        r & 637.13 & 6.97 & 0.37 & 0.17 & 0.07 & 0.0034 & 0.42 \\
        i & 777.42 & 8.66 & 0.45 & 0.26 & 0.12 & 0.0065 & 0.53 \\
        z & 915.79 & 10.16 & 0.54 & 0.35 & 0.27 & 0.0076 & 0.7 \\
        Y & 988.63 & 13.78 & 0.78 & 0.74 & 2.08 & 0.0153 & 2.35 \\\hline
    \end{tabular}
    \caption{IGL values extrapolated from fit of measured galaxy brightness versus source mask dilation size. The total error shown in the right-most column is calculated as the quadrature sum of the error terms derived from the stars-masked and stars-and-galaxies-masked stacked images ($E_*$ and $E_{*G}$), the IGL fitting ($E_{fit}$), and the jackknife analysis ($E_{jack}$). All IGL and error values are presented in ${\rm nW/m^2/sr}$. 
    \label{tab:igl}}
\end{table*}

\section{Discussion and Conclusions}
\begin{figure*}[ht!]
\includegraphics[width=\textwidth]{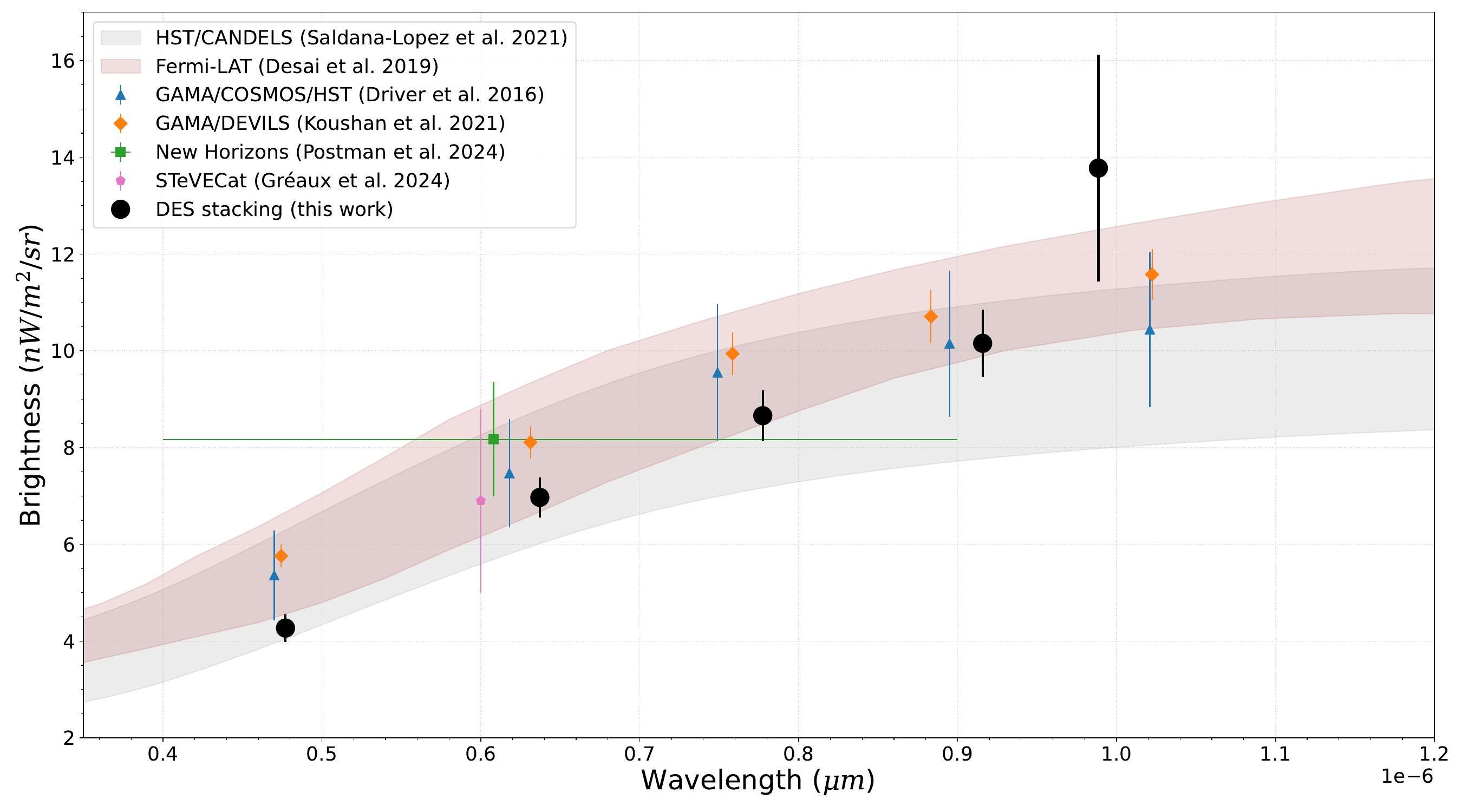}
\caption{Measured IGL values from this work (black circles) compared to previously reported measurements from HST/CANDELS (gray shaded region) \citep{saldanalopez24}, Fermi-LAT (brown shaded region) \citep{desai19}, GAMA/COSMOS/HST (blue triangles) \citep{driver16}, GAMA/DEVILS (orange rhombi) \citep{koushan21}, New Horizons (green square) \citep{postman24}, and STeVECat (pink pentagon) \citep{greaux24}.  
\label{fig:igl_comparison}}
\end{figure*}

Extrapolated IGL values for all five DES bands are given in Table \ref{tab:igl} and shown in Figure \ref{fig:igl_comparison} alongside previously reported estimates from other studies. We find that our measurements are in good agreement with measurements of the IGL derived from galaxy counts and gamma-ray observations, with the exception of the \textit{Y}-band, which is significantly higher than previously reported values and has larger error bars. While each of the individual error terms in Equation \ref{eq:error} tend to increase with increasing wavelength, the \textit{Y}-band in particular has a much larger $E_{fit}$ term relative to the other bands. The \textit{Y}-band IGL fit may benefit from increasing the number of image cutouts used in the stack and/or extending the range of source mask sizes (i.e., to increase the number of points used to calculate the fit).

We have also derived the brightness associated with stars as well as other unclassified sources (Table \ref{tab:brightness}). Roughly, we find the light associated with stars accounts for upwards of two-thirds of the total brightness, while the ``other'' sources account for only $\approx$1\% of the overall brightness. We verified that slight changes to Equation 1 do not appreciably affect the final IGL measurements. While the DES survey is less deep than the ultra-deep HST and JWST surveys used for galaxy counting, our method is complementary to the standard IGL methods, offering greater sky coverage and lower statistical and cosmic variance. Our method accounts for extended source emission by measuring the change in IGL as a function of source mask size and fitting a curve, but may ultimately miss the faintest sources.

In summary, we have presented a new method for measuring the integrated galaxy light (non-diffuse COB) from direct images of the sky. We find our results to be in good agreement with previously reported IGL measurements derived from galaxy counts and gamma-ray observations. Unlike traditional direct-imaging COB techniques, which must precisely model and remove foregrounds such as zodiacal light and diffuse galactic light, this method requires no such corrections. There is a wealth of existing optical and infrared image data from both space and ground-based observatories such as the Wide-field Infrared Survey Explorer (WISE) \citep{wright10}, the Sloan Digital Sky Survey (SDSS) \citep{york00}, and the Panoramic Survey Telescope and Rapid Response System (Pan-STARRS) \citep{chambers16}. In the near future, this will be also possible with wide-field data from the Rubin Observatory \citep{ivezic19} as well as the Spectro-Photometer for the History of the Universe, Epoch of Reionization, and Ices Explorer (SPHEREx) \citep{dore14}. At higher spatial resolutions, and assuming access to the pixels, similar large stacking studies can also be done in the near-IR with the Euclid \citep{euclid24} and Roman \citep{akeson19} space telescopes with their wide fields of view to smooth out over cosmic variance. With the method presented in this work, images from legacy instruments as well as next-generation observatories can be used to carry out additional studies of the IGL at other wavelengths, or in different areas of the sky. 

\begin{acknowledgments}
We would like to thank Skylar Grayson, Patrick Kamienski, Darby Kramer, Rosalia O'Brien, Russell Ryan, Alex van Engelen, and Eve Vavagiakis for useful discussions that greatly improved the manuscript. We also extend our gratitude to the anonymous reviewer, whose suggestions helped improve this paper. ES was supported in part by NASA Grant 80NSSC22K1265. Much of the analysis for this work was carried out on the Sol supercomputer at Arizona State University. 
\end{acknowledgments}

%

\vspace{5mm}


\software{\texttt{Photutils} \citep{bradley22}, \texttt{SourceExtractor} \citep{bertin96}, \texttt{gdpyc}
          }



%


\bibliography{main}{}
\bibliographystyle{aasjournal}



\end{document}